\documentclass[a4paper,fleqn,usenatbib]{mnras}

\usepackage[T1]{fontenc}
\usepackage{ae,aecompl}
\usepackage{url}

\usepackage{graphicx}	
\usepackage{amsmath}	
\usepackage{amssymb}	

\title[X-ray luminosity of Be/X-ray binaries]{Simulating the X-ray luminosity of Be X-ray binaries: the case for
black holes versus neutron stars}

\author[R. O. Brown et al.]{
R. O. Brown,$^{1}$\thanks{E-mail: rob1g10@soton.ac.uk}
W. C. G. Ho,$^{1, 2, 3}$
M. J. Coe,$^{1}$
A. T. Okazaki$^{4}$
\\
$^{1}$Physics and Astronomy, University of Southampton, Southampton, SO17 1BJ, UK \\
$^{2}$Mathematical Sciences and STAG Research Centre, University of Southampton, Southampton, SO17 1BJ, UK  \\
$^{3}$Department of Physics and Astronomy, Haverford College, 370 Lancaster Avenue, Haverford, PA 19041, USA \\
$^{4}$Faculty of Engineering, Hokkai-Gakuen University, Toyohira-ku, Sapporo, 062-8605, Japan
}

\date{Accepted XXX. Received YYY; in original form ZZZ}

\pubyear{2018}

\begin{document}
\label{firstpage}
\pagerange{\pageref{firstpage}--\pageref{lastpage}}
\maketitle

\begin{abstract}
There are over 100 Be stars that are known to have neutron star companions but only one such system with a black hole. Previous theoretical work suggests this is not due to their formation but due to differences in X-ray luminosity. It has also been proposed that the truncation of the Be star's circumstellar disc is dependent on the mass of the compact object. Hence, Be star discs in black hole binaries are smaller. Since accretion onto the compact object from the Be star's disc is what powers the X-ray luminosity, a smaller disc in black hole systems leads to a lower luminosity. In this paper, simulations are performed with a range of eccentricities and compact object mass. The disc's size and density are shown to be dependent on both quantities. Mass capture and, in turn, X-ray luminosity are heavily dependent on the size and density of the disc. Be/black hole binaries are expected to be up to $\sim$10 times fainter than Be/neutron star binaries when both systems have the same eccentricity and can be 100 times fainter when comparing systems with different eccentricity.
\end{abstract}




\section{Introduction}


Be stars are B spectral type stars that are non-supergiant and have, or have had at some time, one or more Balmer lines in emission \citep{JascEgre1982}. In comparison to many other main sequence stars, Be stars have high rates of mass outflow, and rotate very quickly \citep{Slettebak1988}. This rapid rotation, in addition to non-radial pulsations, is thought to lead to a diffuse and gaseous circumstellar disc \citep{Riv2000}. This is commonly referred to as a decretion disc. The disc is a dominant mechanism for accretion onto a binary companion, because the stellar wind of a Be star is generally weak. When the companion is a compact object, the system is a High Mass X-ray Binary (HMXB).

Be/X-ray binaries are the largest population of observable HMXBs \citep{RapHeu1982, HeuRap1987, Coleiro2013}. The varying size of the disc, coupled with the interaction of a compact object leads to a variety of observable effects. Some of these, such as superorbital modulation and giant outbursts, are not well understood. Comprehending these phenomena can lead to a better understanding of the extreme physics of neutron stars and black holes (see \citet{Reig2013} for review).

\subsection{The lack of known Be/black hole binaries} \label{lackBHs}

Binary evolution models have predicted the existence of Be-black hole X-ray binaries, with \citet{RagLip1999} proposing the detection of such systems soon after their findings. They used Monte Carlo simulations to determine that there is an evolutionary track that leads to formation of Be/black hole binaries. They tracked the evolution of 10$^{6}$ zero age main sequence binary systems with initial masses 10 to 120$M_{\odot}$. A logarithmic distribution for initial binary separation was used. The simulations were run with and without kick velocity. Systems that evolved had a wide range of both orbital period and eccentricity. Orbital period varies from less than ten to thousands of days with a peak at tens of days in every dataset. Eccentricity can have any value between 0 and 1 with a maximum between 0.2 and 0.4 for all the simulations. One black hole binary system for every 20-30 neutron star systems was estimated. 

More recent stellar population synthesis models by \citet{ZioBel2011} also used varying kick velocities but modelled common envelope evolution and non-conservative mass transfer via Roche lobe overflow. A solar metallicity ($Z=0.02$) for galactic binaries and a low metallicity ($Z=0.008$) for Magellanic Clouds binaries were adopted. The Milky Way and Magellanic Clouds also had differing stellar initial mass functions. These initial conditions lead to a ratio in the galaxy of 30-50 neutron stars per black hole and a ratio in the Magellanic Clouds of $\sim$10. Therefore the expected number of Be X-ray binaries with black hole companions in the Galaxy should be $\sim$0 - 2 \citep{BelZio2009} and the lack of Be/black hole binaries is not surprising. However, in the Magellanic Clouds, there should be $\sim$6 of these systems.

\citet{ZhangLiWang2004} proposed a greater truncation of the circumstellar disc in systems with shorter orbital periods. Given that some synthesis models predict black hole binaries will form with binary periods of less than $\sim$30 days \citep{Pod2003}, it was proposed that all black holes truncate the Be star disc more effectively than neutron star companions. A smaller disc means interaction and accretion is less likely. Hence all black hole systems would be fainter X-ray objects.


The first confirmed Be-black hole binary system, MWC 656, was found by \citet{Casares2014}. Its luminosity is less than 1.6$\times$10$^{-7}$ times the Eddington luminosity implying extremely inefficient accretion. This further enforces the link between the lack of known black hole binaries and low X-ray luminosity. \citet{Ribo2017} have shown that it has become $\sim$7 times fainter since it was observed in 2014. They also found that the observable quiescent behaviour is fully compatible with low mass X-ray binaries (LMXBs) with black hole companions. This suggests that black hole accretion is independent of the donor star. \citet{Casares2014} state that the Be star has a range of 10-16 $M_{\odot}$. Their solution yields a mass ratio of 0.41$\pm$0.07.  It has an orbital period of 60.37 days and an eccentricity of 0.1. Therefore, it has a periastron distance of $\sim$28$R_{\odot}$. From archival data of ROSAT over 7-11 July 1993 and Swift on 8 March 2011, they set an upper limit on the X-ray luminosity of $L_{X} <$ 10$^{32}$erg s$^{-1}$. MWC 656 will be used as a comparison in the data. Assuming the Be star has the same mass as the simulations in this paper (13$M_{\odot}$), the black hole mass is 5.3$M_{\odot}$. 

\subsection{The distribution of eccentricity in Be/neutron star systems}

\begin{figure}
	\centering
	\includegraphics[width=.5\textwidth]{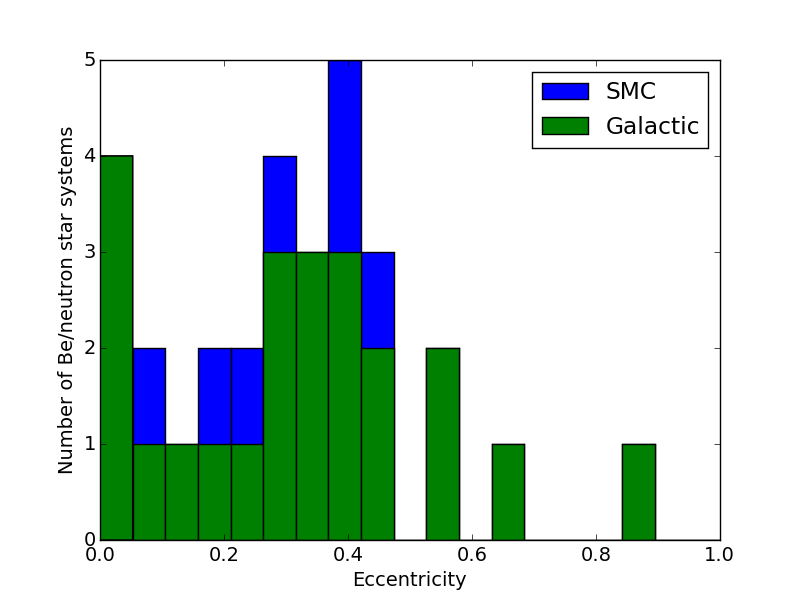}
	\caption{A histogram showing the distribution of Be/neutron star binaries with known eccentricity in the Milky Way and SMC. The bars for the SMC are placed on top of those for the Milky Way. Therefore each bin shows the total number of systems with the given range of eccentricities. Three Be/neutron binary systems have a lower limit on eccentricity of $e > 0.5$ and thus are not included. The values used for this figure are contained in Table \ref{tab:RealSystems}.}
	\label{fig:RealEccentricities}
\end{figure}

Given the importance of eccentricity in accretion in interacting binary systems, it is pertinent to discuss the eccentricity of observed Be/X-ray binaries. Be/neutron star binaries with known eccentricities in the Milky Way, Large Magellanic Cloud (LMC) and Small Magellanic Cloud (SMC) are shown in Table \ref{tab:RealSystems}. Orbital period and maximum observed X-ray luminosity, L$_{\mathrm{max}}$, are also included. Figure \ref{fig:RealEccentricities} illustrates the distribution of these eccentricities. Almost half of Be/neutron star systems have an eccentricity between 0.25 and 0.45. The second largest population is systems with circular orbits, the significance of which is shown in \citet{Pfhal2002}.  In this sample, $\sim$82$\%$ of Be/neutron star binaries have an eccentricity of $e<0.5$. 

\setcounter{table}{0}
\begin{table} 
	\centering
	\caption{Table of eccentricity and maximum observed X-ray luminosity for Be/X-ray systems in the Milky Way, LMC and SMC. (1) \url{http://xray.sai.msu.ru/~raguzova/BeXcat/title.html}. (2) \citet{ReigNes2013} (3) \citet{Reig2016}. (4) \citet{Liu2006}. (5) \citet{CoeKirk2015}. (6) \citet{Townsend2011}. (7) \citet{Coe2015}. (8) \citet{Townsend2011a}. (9) \citet{Schurch2008}. This table is for demonstrating the low X-ray luminosity of MWC 656 and therefore some values may not be the most recent ones.}
	\begin{tabular}{lccc}

  		\hline
		Milky Way & e & L$_{\mathrm{max}}$ (erg s$^{-1}$) & Source \\
		\hline

		$\gamma$ Cas & 0.26 & 3.9 $\times$ 10$^{34}$ & 1\\
		V635 Cas & 0.34 & 3.0 $\times$ 10$^{37}$ & 1 \\
		V615 Cas & 0.537 & 2.0 $\times$ 10$^{34}$ & 1 \\
		X Per & 0.11 & 3.0 $\times$ 10$^{35}$ & 1 \\
		V725 Tau & 0.47 & 2.0 $\times$ 10$^{37}$ & 1 \\
		GS 0834-430 & 0.12 & 1.1 $\times$ 10$^{37}$ & 1 \\ 
		GRO J1008-57 & 0.68 & 2.0 $\times$ 10$^{37}$ & 1, 2 \\
		V801 Cen & >0.5 & 7.4 $\times$ 10$^{34}$ & 1 \\
		GX 304-1 & >0.5 & 1.0 $\times$ 10$^{36}$ & 1 \\ 
		2S 1417-624 & 0.446 & 8.0 $\times$ 10$^{36}$ & 1 \\
		XTE J1543-568 & <0.03 & >10$^{37}$ & 1 \\
		2S 1553-542 & <0.09 & 7.0 $\times$ 10$^{36}$ & 1 \\
		2S 1845-024 & 0.88 & 6.0 $\times$ 10$^{36}$ & 1 \\
		4U 1901+03 & 0.036 & 1.1 $\times$ 10$^{38}$ & 1 \\
		XTE J1946+274 & 0.33 & 5.4 $\times$ 10$^{36}$ & 1 \\
		GRO J1948+32 & 0.03 & 2.1 $\times$ 10$^{37}$ & 1 \\
		EXO 2030+375 & 0.41 & 1.0 $\times$ 10$^{38}$ & 1 \\
		SAX J2103.5+4545 & 0.40 & 3.0 $\times$ 10$^{36}$ & 1 \\ 
		2S 0114+65 & 0.16 & 4.9 $\times$ 10$^{35}$ & 3, 4 \\
		V 0332+53 & 0.42 & 3.4 $\times$ 10$^{38}$ & 2, 3 \\
		AX J1845.0-0433 & 0.34 & 1.4 $\times$ 10$^{36}$ & 3, 4 \\
		4U 1907+09 & 0.28 & 2.3 $\times$ 10$^{37}$ & 3, 4 \\
		KS 1947+300 & 0.03 & 8.8 $\times$ 10$^{37}$ & 3, 4 \\
		4U 2206+54 & 0.30 & 8.8 $\times$ 10$^{34}$ & 3, 4 \\
		MWC 656 (BH) & 0.1 & 3.7 $\times$ 10$^{31}$ & 1 \\

		\hline
		LMC & e & L$_{\mathrm{max}}$ (erg s$^{-1}$) & Source \\
		\hline

		0535-668 & >0.5 & 1.0 $\times$ 10$^{39}$ & 1 \\

		\hline
		SMC & e & L$_{\mathrm{max}}$ (erg s$^{-1}$) & Source \\
		\hline

		SXP 2.37 & 0.07 & 2.1 $\times$ 10$^{38}$ & 5, 6 \\
		SXP 5.05 & 0.16 & 5 $\times$ 10$^{37}$ & 5, 7 \\
        SXP 6.85 & 0.26 & 3.3 $\times$ 10$^{37}$ & 5, 6 \\
        SXP 8.80 & 0.41 & 7.3 $\times$ 10$^{37}$ & 5, 6 \\
        SXP 11.5 & 0.28 & 1 $\times$ 10$^{37}$ & 5, 8 \\
        SXP 18.3 & 0.43 & 3 $\times$ 10$^{37}$ & 5, 9 \\
        SXP 74.7 & 0.4 & 3.5 $\times$ 10$^{37}$ & 5, 6 \\

		\hline

	\end{tabular}
	\label{tab:RealSystems}
\end{table}

\begin{figure}
	\centering
	\includegraphics[width=.5\textwidth]{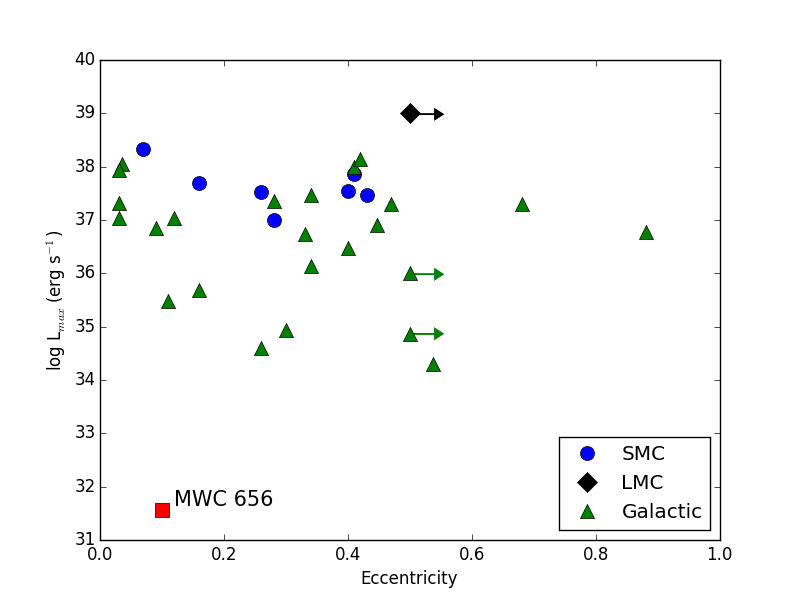}
	\caption{The relationship between maximum observed X-ray luminosity, L$_{max}$, and eccentricity for Be/X-ray binaries in the Milky Way, LMC and SMC. The only Be/black hole system, MWC 656, is also included. Arrows demonstrate the lower limit on eccentricity for the 3 systems with $e>0.5$. The values used for this figure are contained in Table \ref{tab:RealSystems}.}
	\label{fig:RealEccentricitiesVSLmax}
\end{figure}

The relationship between eccentricity and maximum X-ray luminosity is shown in Figure \ref{fig:RealEccentricitiesVSLmax}. The single confirmed Be/black hole binary, MWC 656, is included in addition to the Be/neutron star binaries. There is no clear relation between maximum luminosity and eccentricity for the Be/neutron star binaries. As mentioned in Section \ref{lackBHs}, MWC 656 has a considerably lower maximum X-ray luminosity than all the Be/neutron star binaries and is more than two orders of magnitude fainter than the faintest Be/neutron star system in this sample. 

In this paper, simulations are used to determine the reaction of the Be star's decretion disc to compact objects of different mass. The results predict that Be/black hole binaries are fainter X-ray sources than Be/neutron star binaries. In Section \ref{sec:Simulations}, results from Smoothed Particle Hydrodynamics simulations of Be/X-ray binaries are shown. Systems of varying compact object mass and eccentricity are investigated. The effects on base gas density, disc size and X-ray luminosity are discussed. Section \ref{sec:Conclusions}, discusses the outcome of the results shown in the paper. Results from previous work are considered and predictions are made for the possibility of detections in the future.


\section{Simulations} \label{sec:Simulations}

A 3D smoothed particle hydrodynamics (SPH) code is used to simulate Be/X-ray binaries and the results are discussed in this section. The Be star's decretion disc is modelled by an ensemble of particles each of mass $\sim 10^{-15}$M$_{\odot}$. The disc is assumed to be isothermal for simplicity. The particles are injected into the disc with Keplerian velocity at a random azimuthal angle at 1.05 stellar radii from the centre of the Be star. They are placed at a random small distance from the equatorial plane. The Be star and compact object are modelled using sink particles. For further details on the code, see \citet{Bate1995}, \citet{Okazaki2002} and \citet{Hayasaki2004}.

All systems were evolved until they reached equilibrium. If the number of particles in the disc around an orbit changes by less than 1$\%$ for more than 5 orbits, the system is considered to be in a equilibrium. The number of particles in the disc at equilibrium ranges from $\sim$20,000 to $\sim$200,000. This is dependent of the orbital period of the system.

\subsection{Properties of simulated systems}

\begin{figure*}
	\centering
	\includegraphics[width=1.05\textwidth]{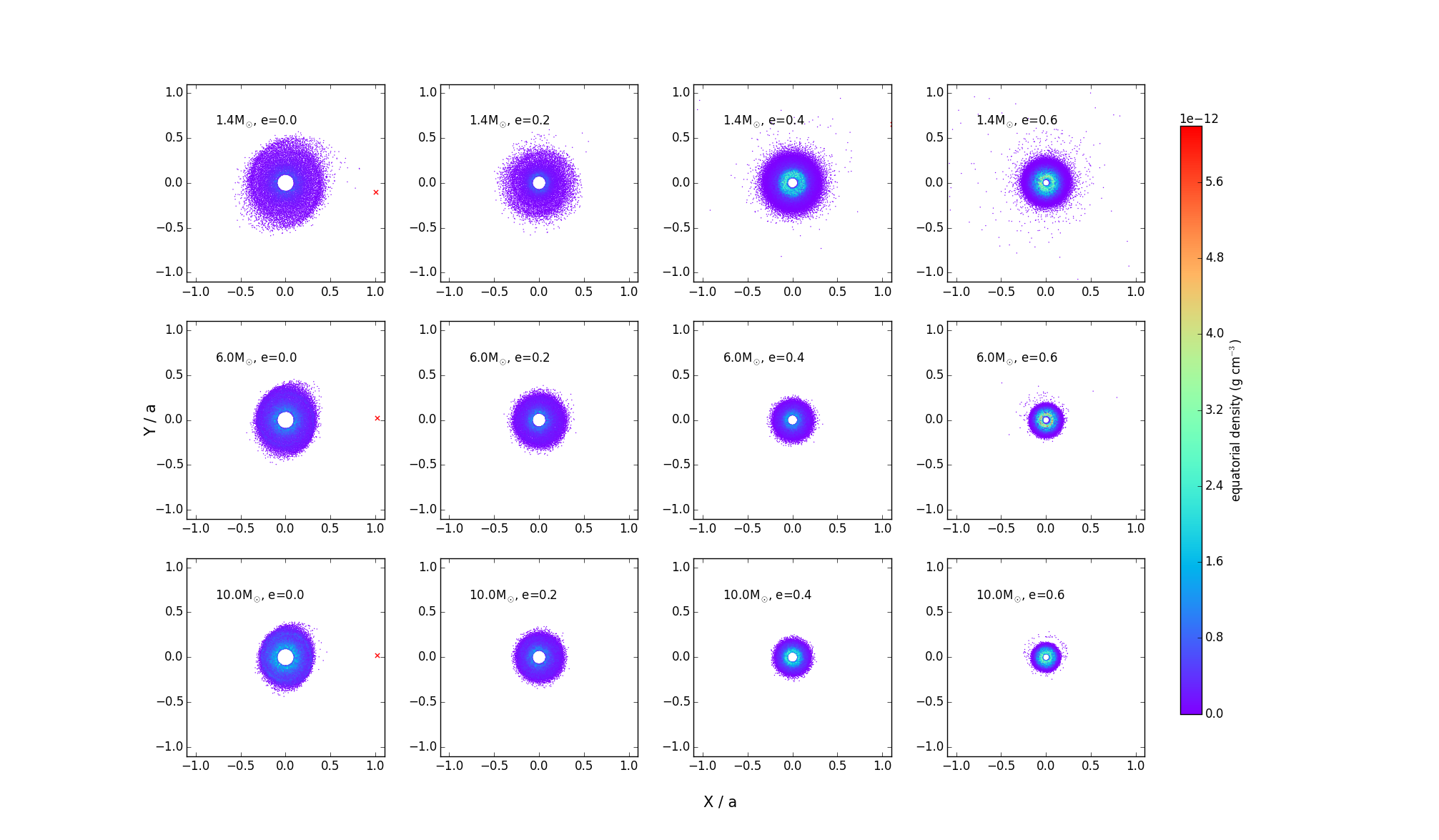}
	\includegraphics[width=1.05\textwidth]{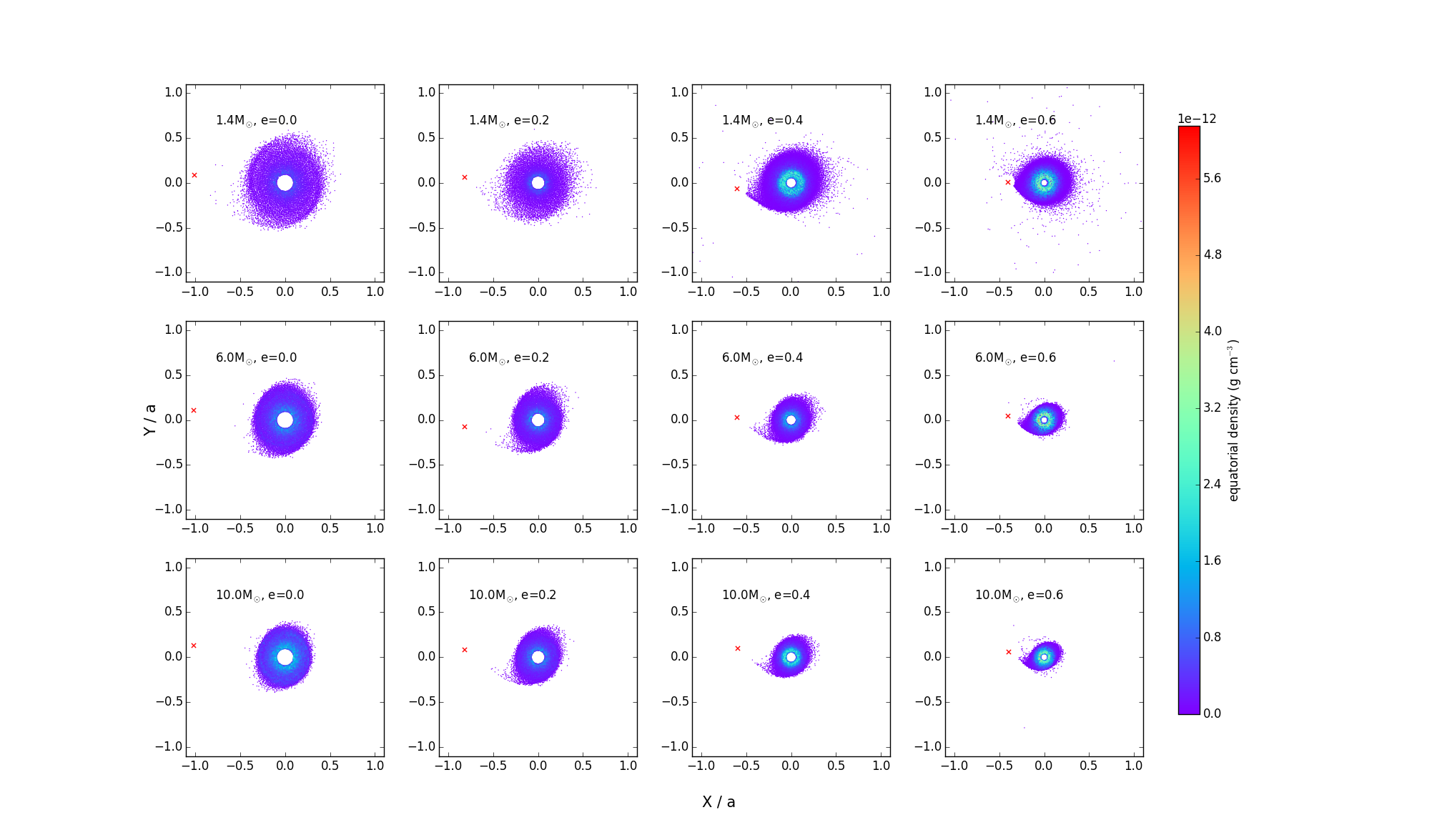}
	\caption{Images of the circumstellar disc normalised by the semi-major axis, a. Three different compact object masses are shown (increasing from top to bottom) and four different eccentricities (increasing from left to right). The parameters of the systems shown are M=1.4$M_{\odot}$, 6$M_{\odot}$, 10$M_{\odot}$ and e=0.0, 0.2, 0.4 and 0.6. The colorbar shows the equatorial density within the disc. The compact object is marked with a red cross and lies close to apastron (top 4x3 grid) or periastron (bottom 4x3 grid). Periastron distance is the same for all the simulations.}
	\label{fig:discSize}
\end{figure*}

All systems have a Be star of mass 13$M_{\odot}$ and radius 7$R_{\odot}$, to match a main sequence star with a B1V spectral type \citep{Silaj2014}. The expected range of mass ejection for Be stars is $10^{-12} - 10^{-9} M_{\odot}$yr$^{-1}$ \citep{Vieira2015}. A constant mass ejection of $\sim$10$^{-11}M_{\odot}$yr$^{-1}$ is assumed.   

All orbits are coplanar and the periastron distance is kept at $\sim$80 solar radii because the truncation of the disc is heavily dependent on it. It is the semi-major axis for a neutron star in a circular orbit with an orbital period of 20 days. The orbital period of every simulation is shown in Table \ref{tab:Porb}.

\setcounter{table}{1}
\begin{table}
	\centering
	\caption{Table of assumed orbital periods for all neutron star and 10 M$_\odot$ black hole systems used as input for our simulations. Compact object mass, M$_{X}$, is in solar masses. Note that for MWC 656, M$_{X}$ = 5.3 M$_{\odot}$, P$_{orb}$=60.37 and $e=0.1$, assuming a Be star mass of 13 M$_{\odot}$.}
    \label{tab:Porb}
	\begin{tabular}{lcc} 
		\hline
		$e$ & P$_{orb}$ (M$_{X}$=1.4 M$_\odot$) & P$_{orb}$ (M$_{X}$=10 M$_\odot$)  \\
		\hline
		0.0 & 20   days & 15.8 days \\
		0.1 & 23.4 days & 18.5 days \\
		0.2 & 28   days & 22.1 days \\
		0.3 & 34.1 days & 27   days \\
		0.4 & 43   days & 34.1 days \\
		0.5 & 56.6 days & 44.8 days \\
		0.6 & 79.1 days & 62.6 days \\
		\hline
	\end{tabular}
\end{table}

The density profile of a Be star's decretion disc is well modelled by the following equation \citep{TouhGies2011}

\begin{equation} \label{eq:TouhGies}
\rho(r, z) = \rho_{0} \left(\frac{r}{R_{*}}\right)^{-n} \exp \left[ -\frac{1}{2}\left(\frac{z}{H(r)}\right)^{2}\right],
\end{equation}

\noindent where $\rho_{0}$ is the base gas density, $n$ is the radial density exponent and $r$ and $z$ are the radial and vertical cylindrical coordinates, respectively. The power law coefficient, $n$, is variable between 2.5 and 4 \citep{RivCarc2013}. $H(r)$ is the disc's vertical scale height and is given by

\begin{equation} \label{eq:scaleHeight}
H(r) = \frac{c_{s}}{V_{K}} \left(\frac{r}{R_{*}}\right)^{\frac{3}{2}},
\end{equation}

\noindent where $V_{K}$ is the Keplerian velocity at the stellar equator and $c_{s}$ is the speed of sound in the disc, which is temperature dependent.

The base gas density of the disc can approximate the overall density profile as shown by Equation \ref{eq:TouhGies}. For eccentricities of $e \leq 0.1$, base gas density increases with compact object mass. The disc is compacted by the black hole or neutron star that spends all of its time in close proximity to the disc. For higher eccentricities, it has the opposite relationship. These systems have longer orbital periods and the compact object spends a smaller amount of time near the decretion disc. This is less disruptive and allows the disc to increase in size and decrease in density.






\subsection{Truncation of the disc} \label{sec:truncation}

\begin{figure}
	\centering
	\includegraphics[width=.5\textwidth]{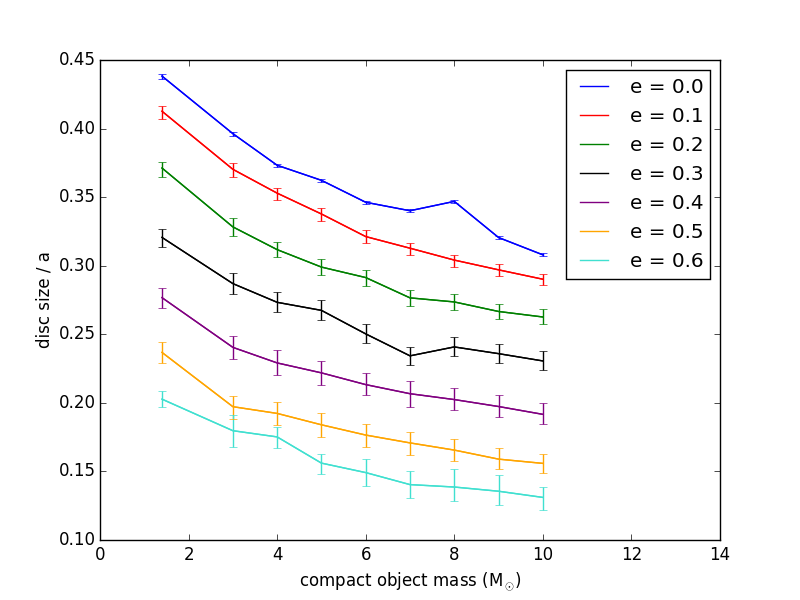}
	\caption{The size of the Be star's decretion disc normalised by the semi-major axis, a, for all systems. Disc size is taken to be the radius within which 90$\%$ of all the simulations particles are contained. Periastron distance is the same for all the simulations.}
	\label{fig:discSizes}
\end{figure}

The truncation of the Be star's disc is an established feature of Be/X-ray binaries \citep{Stefl2007}. Figure \ref{fig:discSize} illustrates the size of the disc for 12 of the systems presented in this paper and Figure \ref{fig:discSizes} shows the numerical relationship between disc size and compact object mass. The size of the disc is defined as the radius that contains 90$\%$ of the simulation particles. The length scale has been normalised by the semi-major axis, $a$. This removes dependence on the size of the orbit. The Be star's disc is truncated at smaller (non-dimensional) radii for higher orbital eccentricities. 

\begin{figure}
	\centering
	\includegraphics[width=.5\textwidth]{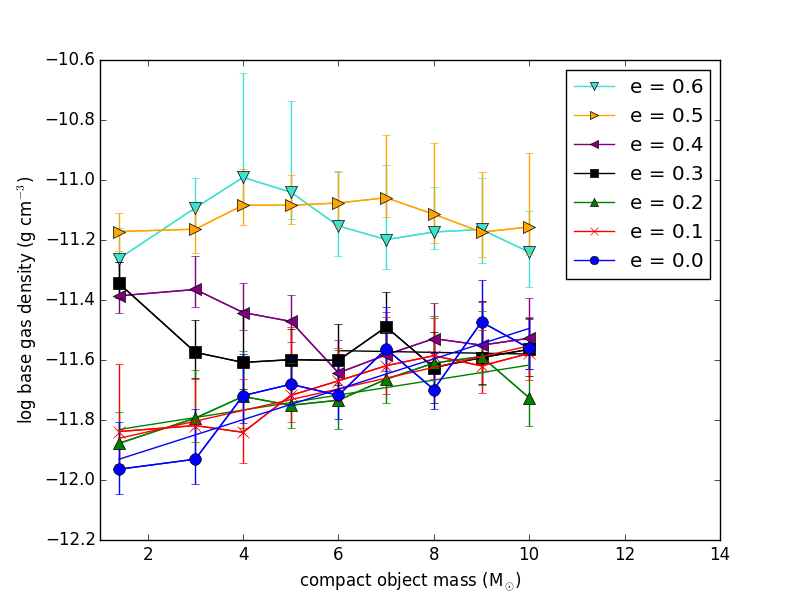}
	\caption{Density of the disc as a function of compact object mass. Vertical bars show the range of base gas density around an orbit, and points show the average over the orbit.}
	\label{fig:densities}
\end{figure}

Figure \ref{fig:densities} shows the base gas density for all systems. For the simulations of $e \le 0.2$, base gas density increases with compact object mass. Base gas density decreases by less than a factor of two for $0.2 < e < 0.4$. For $e > 0.4$, base gas density is almost constant.


\subsection{X-ray luminosity}\label{sec:XrayLuminosity}

The relative difference in X-ray luminosity between Be/black hole binaries and Be/neutron star binaries can be estimated using the mass capture of the compact object. For simplicity, it is assumed that all particles captured by the compact object are accreted instantaneously. This yields X-ray luminosities per unit time. X-ray luminosity is calculated using

\begin{equation} 
L_{X} = \frac{G M_{X} \dot{M}} {R_{X}}
\label{eq:LX}
\end{equation}

\noindent for neutron star binaries and

\begin{equation} 
L_{X} = \eta \dot{M} c^{2}
\label{eq:LX}
\end{equation}

\noindent for black hole binaries. Here, $\dot{M}$ is the rate of accretion, $M_{X}$ and $R_{X}$ are the compact object's mass and radius respectively and $\eta$ is the conversion efficiency of the rest mass energy of accreted matter into radiation, where $\eta=0.1$ is adopted in this paper \citep{accBook}. The number of captured particles is converted into solar masses using $M / M_{\odot} = 2.867 \times 10^{-15} \times N$, where $N$ is the number of particles.

\begin{figure}
	\centering
	\includegraphics[width=.5\textwidth]{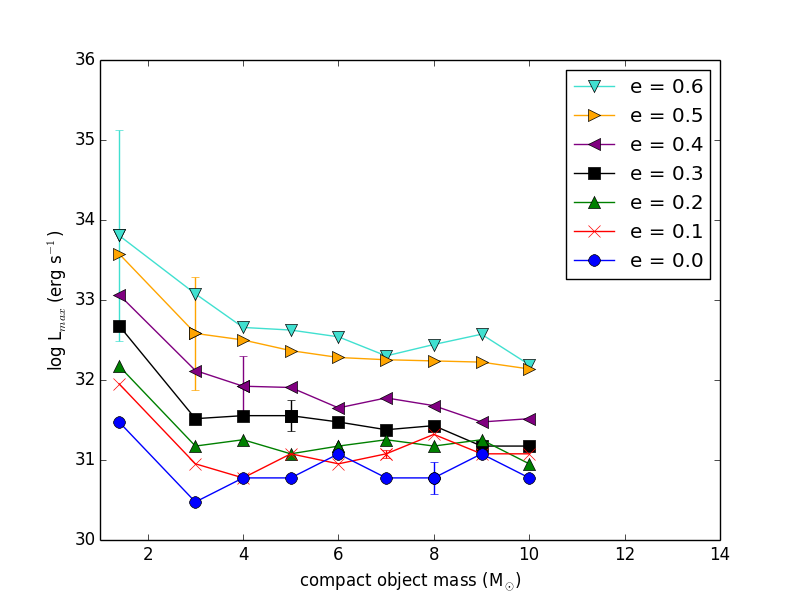}
	\caption{The peak daily X-ray luminosity around an orbit. Errorbars show $\sqrt{N}$ where $N$ is the number of particles captured and converted into an X-ray luminosity (see text).}
	\label{fig:Lmax}
\end{figure}

\begin{figure}
	\centering
	\includegraphics[width=.5\textwidth]{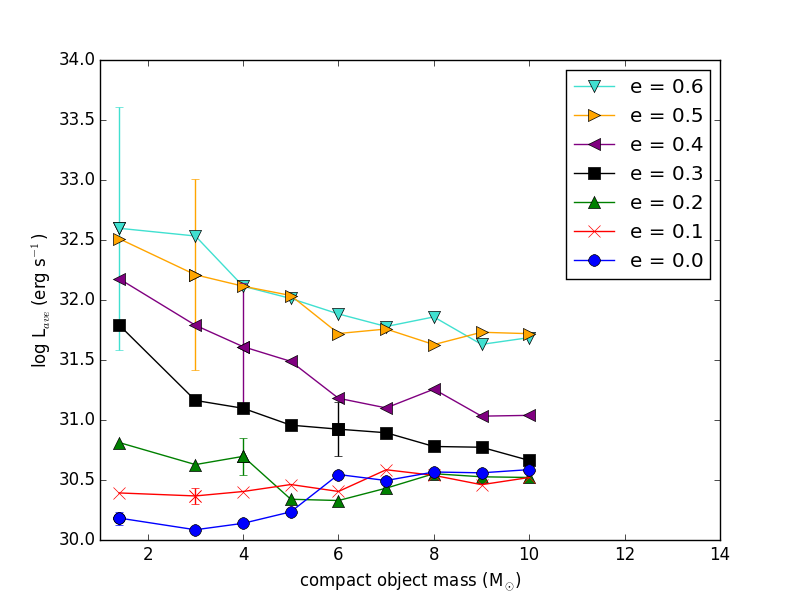}
	\caption{The average daily X-ray luminosity around an orbit. Errorbars show $\sqrt{N}$ where $N$ is the number of particles captured and converted into an X-ray luminosity (see text).}
	\label{fig:Lave}
\end{figure}

The maximum measured value of X-ray luminosity over 5 orbits is defined as the peak luminosity. The average luminosity is the median value of the same 5 orbits. Figures \ref{fig:Lmax} and \ref{fig:Lave} show the peak and average X-ray luminosity for all simulations. The bars on the plots show $\sqrt{N}$ where $N$ is the number of captured particles converted into a luminosity. For the binaries with $e < 0.2$, X-ray luminosity falls with increasing compact object mass. For systems of $e \le 0.2$, the number of captured particles is small. Thus, although X-ray luminosity varies with compact object mass, the variations are up to 5000 times smaller than those for other eccentricities. Peak and average luminosity have a similar dependence on compact object mass. There is up to an order of magnitude difference between the peak luminosity of a neutron star and a 10M$_{\odot}$ black hole simulation of equal eccentricity. When comparing simulations of differing eccentricity, this disparity is three orders of magnitude. 


\subsection{Binary phase of accretion}

\begin{figure}
	\centering
	\includegraphics[width=.5\textwidth]{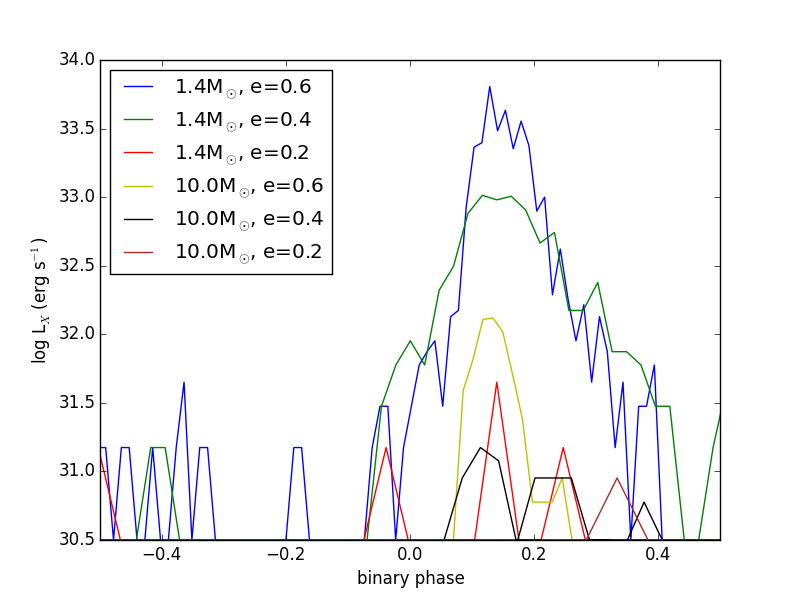}
	\caption{The logarithm of X-ray luminosity around an orbit for the neutron star and the 10$M_{\odot}$ black hole with eccentricities $e = 0.2, 0.4,$ and $0.6$. Periastron is at a binary phase of 0.0. The peak mass capture for all systems lies at a binary phase $\sim$0.15. }
	\label{fig:LOGbphase}
\end{figure}

It is often assumed that the maximum X-ray luminosity occurs at periastron. However, there may be a delay between these two events. Previous work on neutron star mass capture by \citet{Okazaki2004} shows there is a delay between periastron and peak mass capture. Figure \ref{fig:LOGbphase} shows the relation between X-ray luminosity and binary phase. The peak X-ray luminosity occurs at a binary phase of $\sim$0.15. Note that accretion onto the compact object is not modelled accurately, and therefore, a precise estimate of timing of peak X-ray luminosity cannot be made. Nevertheless, this result confirms the delay found in previous work and reveals that there is no dependence on compact object mass. 

Increasing eccentricity results in higher X-ray luminosities over shorter ranges of orbital phase. This can be understood from the discussion of truncation in Section \ref{sec:truncation}. The smaller gap between the truncated disc and periastron in systems of higher eccentricity leads to a higher rate of mass capture. Because more massive compact objects produce more significant truncation of the decretion disc, there is a shorter period during which accretion takes place.

\section{Modelling MWC 656}

The maximum observed X-ray luminosity of MWC 656 is 500 times smaller than the faintest Be/neutron star binary in the sample shown in Table \ref{tab:RealSystems} and Figure \ref{fig:RealEccentricitiesVSLmax}. It is almost 10,000 times fainter than Be/neutron star binaries of similar eccentricity. We see from Figure \ref{fig:Lmax} that, for a given eccentricity, a change in compact object mass can only produce a change in L$_{\mathrm{max}}$ of $\sim$20-30. 

\begin{table}
	\centering
	\caption{Properties of the MWC 656 simulation. The table contains the base gas density, the disc size normalised by the semi-major axis, the peak daily X-ray luminosity and the average daily X-ray luminosity.}
    \label{tab:MWC656}
	\begin{tabular}{lc} 
		\hline
		Disc size / a & 0.28 $\pm$ 0.1 \\
		$\rho_{0}$ (g$\hspace{0.4mm}$cm$^{-3}$) & (8.4 $\pm$ 1.4) $\times$ 10$^{-12}$ \\
		L$_{\mathrm{max}}$ (erg s$^{-1}$) & (3.0 $\pm$ 0.7) $\times$ 10$^{32}$ \\
		L$_{\mathrm{ave}}$ (erg s$^{-1}$) & (5.1 $\pm$ 0.6) $\times$ 10$^{31}$ \\
		\hline
	\end{tabular}
\end{table}

A simulation of MWC 656 was performed to use as a comparison between the simulations and the sample data. Note that the periastron distance of MWC 656 is $\sim$2 times larger than the periastron distance for simulations shown in Section \ref{sec:Simulations}. The base gas density, normalised disc size, peak and average X-ray luminosity are shown in Table \ref{tab:MWC656}. 

The base gas density is comparable to the highest values seen in the other simulations. The normalised size of the disc is smaller than a 5M$_{\odot}$ black hole with an eccentricity of 0.1. Therefore, a larger periastron distance creates a denser decretion disc and normalised disc size does not remain constant when varying periastron distance. In dimensional units, however, the disc of MWC 656 is still larger. Both peak and average X-ray luminosity are higher than a 5M$_{\odot}$ black hole with an eccentricity of 0.1. X-ray luminosity increases with periastron distance when using constant mass ejection rate. 

Further evidence for the inefficient accretion suggested by \citet{Casares2014} is shown by this simulation data. The peak luminosity is greater than the observed maximum for MWC 656. This is despite the mass ejection being less than average for a Be star. A larger periastron distance and, in turn, a less efficient truncation of the disc, leads to a denser disc which allows the accretion of more matter onto the compact object.

This simulation of MWC 656 still has a lower X-ray luminosity than the majority of the other simulated accreting neutron stars. However, the results cannot explain the huge difference between the observable X-ray luminosity between MWC 656 and other observable neutron star systems. This implies the need for some other explanation as to why MWC 656 is extremely faint. A difference in accretion efficiency is the most likely explanation. While the standard type of accretion was assumed in this paper, a Radiatively Inefficient Accretion Flows, or RIAF, is likely to occur for such a low accretion rate regime, where the efficiency of accretion has been shown to be as low as $\eta \sim 0.0045$ \citep{Abram2001}. The X-ray luminosity from a RIAF with the simulated accretion rate for MWC 656 would therefore be $\sim 10^{30}$, which leads to a similar discrepancy between the X-ray luminosity of MWC 656 and the brightest neutron stars for the simulations and observations.







\section{Conclusions} \label{sec:Conclusions}

In this paper, a possible difference in X-ray luminosity between Be/neutron star and Be/black hole binaries was investigated. Simulations of Be/X-ray binaries with varying compact object mass were performed. It is shown that disc size, base gas density and X-ray luminosity are dependent on the mass of the compact object.

The normalised disc size decreases by up to a factor of 1.5 with compact object mass, confirming the more efficient truncation of the Be star disc black holes in Be/X-ray binaries. Base gas density increases with compact object mass by a factor of $\sim$2.5 for systems of $e \le 0.2$. For systems of $0.2 < e < 0.4$ variations are of the same order but decrease with compact object mass. Above $e = 0.4$, the relation is closer to constant.


The simulations discussed above predict that the most luminous X-ray systems are ones with high eccentricity and a Be star circumstellar disc that extends to or outside the periastron distance. However, black hole binaries that fulfill both these criteria may still be fainter than neutron star systems that satisfy only one. 

With the results in this paper and in the absence of X-ray pulsations, it is extremely difficult to distinguish between a black hole and a neutron star in a Be/X-ray binary with X-ray observations alone. However, with knowledge of the orbital parameters and the size of the disc it is plausible. Be/X-ray binaries with a 10M$_{\odot}$ black hole can be more than 10 times fainter than a neutron star system with the same accretion efficiency and periastron distance.

\section*{Acknowledgements}

ROB is grateful for the EA Milne Travel Fellowship from the Royal Astronomical Society to visit and work with ATO. ROB also acknowledges support from the Engineering and Physical Sciences Research Council Centre for Doctoral Training grant EP/L015382/1. WCGH acknowledges support from the Science and Technology Facilities Council through grant number ST/M000931/1. We thank the anonymous reviewer for their helpful comments.






\newpage



\appendix




\label{lastpage}
\end{document}